\documentclass[useAMS, usenatbib]{mn2e}
\usepackage{graphicx}
\usepackage{natbib}     
\usepackage{amsmath} 
\usepackage{float} 
\usepackage{amssymb}    
\usepackage{footnote}
\usepackage{longtable}
\title[Classical T-Tauri stars in Sh\,2-012]{Classical T-Tauri stars with VPHAS+: II: NGC\,6383 in Sh\,2-012\thanks{Source code to calculate the photometric H$\alpha$ emission line widths and associated stellar models for the VPHAS+ filters can be found at https://github.com/astroquackers/HaEW}\thanks{Table 1 is available in electronic form only}}
\author[V.~M.~Kalari]{V.~M.~Kalari$^{1,2}$\thanks{E-mail:
:\,venukalari@gmail.com} \\
$^{1}$Gemini Observatory, Southern Operations Center, c/o AURA, Casilla 603, La Serena, Chile\\
$^{2}$Gemini-CONICYT, Departamento de Astronom{\'i}a, Universidad de Chile, Casilla 36-D Santiago, Chile}
\begin{document}
    
\date{16 December 2014}

\pagerange{\pageref{firstpage}--\pageref{lastpage}} \pubyear{2014}

\maketitle

\label{firstpage}

\begin{abstract}

This paper presents optical\,($ugri$H$\alpha$)--infrared\,({\it JHK}s,3.6--8.0$\mu$m) photometry, and {\it {\it Gaia}} astrometry of 55 Classical T-Tauri stars (CTTS) in the star-forming region Sh\,2-012, and it's central cluster NGC\,6383. The sample was identified based on photometric H$\alpha$ emission line widths, and has a median age of 2.8$\pm$1.6\,Myr, with a mass range between 0.3--1\,$M_{\odot}$. 94\% of CTTS with near-infrared cross-matches fall on the near-infrared T-Tauri locus, with all stars having mid-infrared photometry exhibiting evidence for accreting circumstellar discs. CTTS are found concentrated around the central cluster NGC\,6383, and towards the bright rims located at the edges of Sh\,2-012. Stars across the region have similar ages, suggestive of a single burst of star formation. Mass accretion rates ($\dot{M}_{\rmn{acc}}$) estimated via H$\alpha$ and $u$-band line intensities show a scatter (0.3\,dex) similar to spectroscopic studies, indicating the suitability of H$\alpha$ photometry to estimate $\dot{M}_{\rmn{acc}}$. Examining the variation of $\dot{M}_{\rmn{acc}}$ with stellar mass ($M_{\ast}$), we find a smaller intercept in the $\dot{M}_{\rmn{acc}}$-$M_{\ast}$ relation than oft-quoted in the literature, providing evidence to discriminate between competing theories of protoplanetary disc evolution.
\end{abstract}
 

\setlength{\parskip}{0.1 cm plus2mm minus2mm}
\setlength{\textfloatsep}{0.5 cm}  

\section{Introduction}




In the current paradigm of pre-main sequence (PMS) evolution, optically visible, PMS stars, i.e. Classical T-Tauri stars (CTTS) accrete gas from their circumstellar disc via a stellar magnetosphere \citep{konigl, Calhart92}, facilitated in part by removal of angular momentum via outflows. This accretion processes results in a combination of unique signatures including excess ultraviolet continuum and line emission (notably in H$\alpha$); and infrared excess (at $\lambda>$2$\mu$m) due to the circumstellar disc \citep{gullbring98, muz03}. Mass accretion ($\dot{M}_{\rmn{acc}}$) results in mass gain eventually leading to the main-sequence turn on, while the remaining disc coagulates and forms planetary systems. How T-Tauri stars accrete mass from their disc while losing angular momentum remains an outstanding question \citep{Hart16}. Understanding this process is essential as it controls the evolution of protoplanetary discs and the planetary systems within \citep{disc}. Currently, viscous evolution \citep{hart98, balbus} acting in conjunction with sources of ionisation (for e.g. gamma-rays in \cite{gammie}, or as summarised in \cite{gorti, ercolano} from internal photoevaporation,  UV-radiation, or X-rays) are considered the chief accretion drivers. Hydrodynamic turbulence \citep{hart18}, or Bondi-Hoyle accretion \citep{pad06, Hart08} have also been proposed. These competing theories predict variations in how $\dot{M}_{\rmn{acc}}$ of PMS stars changes with respect to stellar mass and age. Hence, obtaining large samples of stellar and accretion estimates of PMS stars is the key to understanding disc evolution.

The observational study of accretion in young stars has been the subject of a large body of observational work (see \citealt{Hart16} for a recent review). Conventionally, accretion properties of PMS stars are obtained from spectroscopic studies of a few select objects (e.g., \citealt{muz03, natta06, herc08, kalarivink, manara17}), either based on their UV continuum excess, or emission line strengths. Recently \cite{de10, geert11}, and \cite{Kalari15} demonstrated a method using H$\alpha$ photometry to estimate $\dot{M}_{\rmn{acc}}$, with the results comparable to emission-line spectroscopy, and $U$-band photometry respectively. This opens the possibility of identifying CTTS and estimating their $\dot{M}_{\rmn{acc}}$ and stellar properties in a homogeneous manner with clear detection limits. Thereby, populating the sample of measured accretion rates as a function of stellar mass and age uniformly. In tandem with current astrometric \citep{gaia} and deep infrared photometric (e.g., \citealt{glimpse, vvv}) surveys to remove possible outliers and identify discs among the PMS population respectively, we are endowed with powerful tools to study the accretion and disc properties of pre-main sequence stars homogeneously in a given star-forming region.

Such work utilising large surveys to estimate accretion properties homogeneously was previously carried out by \cite{roman02, de10, geert11, rig11, manara12} and \cite{Kalari15}. In \cite{Kalari15}, we used $ugri$H$\alpha$ optical photometry from VPHAS+ (VST/OmegaCAM Photometric H-Alpha Survey), along with archival near and mid-infrared data to identify CTTS in the star-forming region Lagoon Nebula to estimate for the first time $\dot{M}_{\rmn{acc}}$ from H$\alpha$ and $u$-band data, and compare our results to spectroscopic measurements (finding an almost 1:1 relation within 2$\sigma$). In this paper, we continue the work started by studying another star-forming region in the Sagittarius OB\,1 association (along with NGC\,6530 studied in \cite{Kalari15} and NGC\,6531) Sh\,2-012. In addition to the optical VPHAS+ and infrared survey photometry, we utilise the recent {\it Gaia} data release (DR2) which provides proper motions, and parallaxes of stars with similar limiting magnitudes as the optical and infrared data. Taken together, these data allow for a valuable insight into the accretion, disc and kinematic properties of CTTS in Sh\,2-012 in a homogeneous manner. 

\subsection{Sh\,2-012 and the central cluster NGC\,6383}

Sh\,2-012 (also known as RCW\,132; or Gum\,67) is a prime target for this study \citep{sharpless}. It is a comparatively poorly studied star-forming H{\scriptsize II} region in the Carina-Sagittarius arm \citep{Rauw08}, and forms part of the larger Sagittarius OB\,1 association of star-forming regions found along this section. The central open cluster NGC\,6383 (also Collinder 335) is surrounded by the larger H{\scriptsize II} region encompassing a total area of around 2 sq. deg. It is bounded by a diffuse shell-shaped structure having a radius of nearly 1$\degr$. Most likely, the diffuse shell is ionised by the central star of NGC\,6383, HD\,159176. An O7\,V binary, HD\,159176, has an age of 2.3--2.8\,Myr \citep{Rauw10}. The diffuse shell is resplendent with numerous bright pillars, and rims, which may be the site of current and future star formation. The region is dotted with small optically dark clouds, indicating regions of possibly higher extinction \citep{Rauw10}. Although approximately circular in shape, the H{\scriptsize II} region is not uniformly ionised.

Majority of the cluster population of NGC\,6383 has a rather low and uniform reddening of $E(B-V)$ of 0.32. NGC\,6383 is conveniently located at a distance of around 1300\,pc (although low and high distances have been reported in the review by \citealt{Rauw08}) for studying the low-mass CTTS population with VPHAS+ photometry; as the dynamic magnitude range of VPHAS+ (13$<r<$21) corresponds to a mass range of 0.2--2\,$M_{\odot}$ at this distance and reddening. Assuming that the central cluster and the surrounding population have similar reddening and distance, Sh\,2-012 is an interesting and fruitful place to observe on-going star formation through the study of CTTS. In the literature, the central cluster has been subject of past studies. Briefly, early optical broadband photoelectric photometry suggested a young age of $\sim$2\,Myr for NGC\,6383 \citep{eggen,fitz, loydevans}. Follow-up optical and near-infrared photometric and spectroscopic studies found that many stars had infrared excesses resembling discs, further providing evidence for the presence of CTTS, and on-going star-formation \citep{the, denancker, paun}. A X-ray study by \cite{rauw02} found evidence for a number of candidate pre-main sequence stars. Finally, a detailed recent study using a combination of optical spectroscopy and $UBV$($RI$)$_c$H$\alpha$ photometry was conducted by \cite{Rauw10} in a large area around the central cluster and covering the H{\scriptsize II} region. The authors in addition to confirming a distance of around 1300\,pc and a mean reddening towards cluster members of 0.32, found that X-ray detected pre-main sequence candidates had low levels of H$\alpha$ emission (which may be likely due to them being Weak Line T-Tauri stars; see e.g. \citealt{Hart16}), and also severe contamination by H$\alpha$ emitters from foreground/background populations. This is not surprising given the close location of the region to the Galactic centre, and it's location in our line of sight towards the Galactic arm. Although no detailed study of the formation scenario has been put forward in the literature, \cite{Rauw08} have suggested that the numerous bright rims and pillars maybe the site of future triggered star formation. 

Therefore, it is interesting to study the CTTS in the region from a perspective of gaining information about pre-main sequence accretion, but also understanding the formation scenario of the region. From a study of the complete H{\scriptsize II} region, we are able to add significantly to the literature by presenting the first detailed study of CTTS in the entire Sh\,2-012 region. As this study will be based on identifying accretors based on H$\alpha$ emission and infrared excesses, it negates the contaminants introduced by studying pre-main sequence stars identified purely from colour-magnitude diagrams. With the addition of kinematic information from the recent {\it Gaia} data release, we can exclude the large number of contaminants found in the H$\alpha$ based study of \cite{Rauw10}. With this information in hand, we can study the properties of a large number of CTTS identified in a homogeneous manner, and also study the detailed formation scenario and substructures in the Sh\,2-012 region. 

This paper is organised as follows-- in Section 2 we present the data utilised in this study. In section 3 we identify CTTS and estimate their stellar, accretion and disc properties. In Section 4, we discuss the star formation scenario in Sh\,2-012, and accretion properties of the identified CTTS. Finally, a summary of our results is presented in Section 5.

\begin{figure}
\center 
\includegraphics[width=86mm, height=58mm]{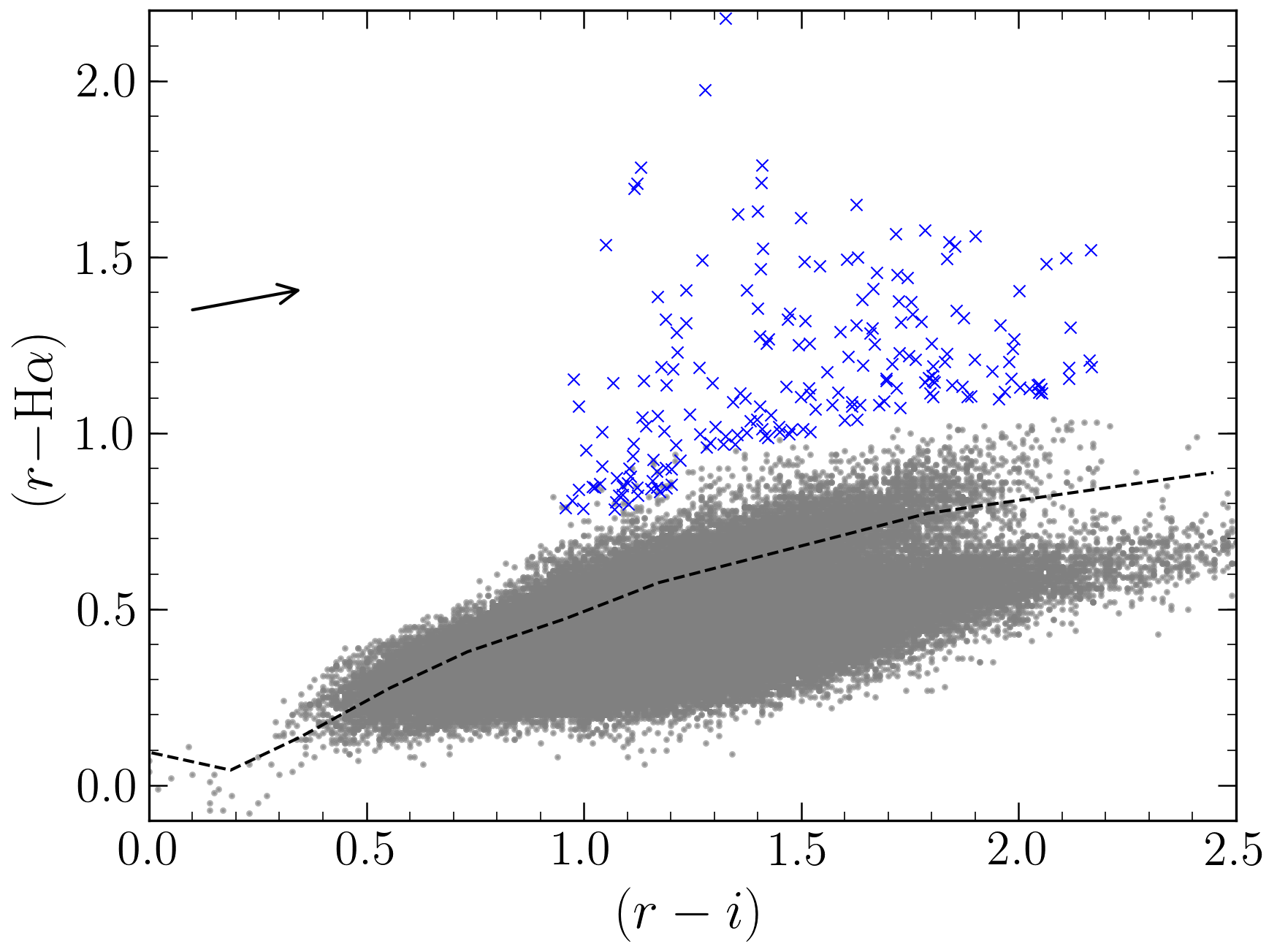}
\center
\caption{($r-$H$\alpha$) vs. ($r-i$) diagram. The dashed line is the interpolated model track reddened by $E(B-V)$\,=\,0.32 assuming a standard Galactic reddening law. Grey dots are stars meeting our selection criterion, and blue crosses are stars meeting our selection criterion that are identified as candidate CTTS based on their EW$_{\rmn{H}\alpha}$. The reddening vector for $A_{V}$\,=\,1 is also shown.}
\label{fig:rha}
\end{figure} 


\section{Data}

\subsection{VPHAS+ imaging}
The VPHAS+ survey \citep{drew14} observed the Southern Galactic Plane and bulge in the $ugri$H$\alpha$ filters using the OmegaCAM CCD imager mounted on the 2.6\,metre VLT Survey Telescope (VST) located at Cerro Paranal, Chile. The OmegaCAM CCD consists of 32\,$\times$\,32 E2V chips which capture a $1^{\circ}$ square field of view at a resolution of 0.21$\arcsec$\,pixel$^{-1}$. Gaps between the chips are minimised by offset exposures, and the total coverage of each field of view is greater than 98\,per\,cent. The $ugri$ band passes are Sloan broadband filters, while the custom-built H$\alpha$ filter has a central wavelength and bandpass of 6588 and 107\,\AA~respectively. Exposure times in these filters are 150, 30, 25, 25, and 120\emph{s} respectively, and observations reach a 5$\sigma$ depth at H$\alpha$\,=\,20.5-21.0\,mag and $g$\,=\,22.2\,-\,22.7\,mag in crowded fields. Practical constraints have meant that the blue ($ug$) and red ($ri$H$\alpha$) observations are carried out separately. Therefore, an additional $r$ observation is carried out with every blue observation to serve as a linking reference.


To identify and characterise CTTS candidates in the star-forming region Sh\,2-012, we define our area of study as a 2$\degr$\,$\times$\,2$\degr$ region centred on Right Ascension (J2000) 17$^{h}$34$^{m}$09$^{s}$, Declination (J2000) $-$32$\degr$32$\arcmin$10$\arcsec$. The data used in this publication were part of the VPHAS+ second data release. The reader is referred to the release document{\footnote{http://www.eso.org/qi/catalogQuery/index/59}} for details of the data reduction procedure. To summarise, the IRAF{\footnote{IRAF is distributed by the National Optical Astronomy Observatory, which is operated by the Association of Universities for Research in Astronomy (AURA) under a cooperative agreement with the National Science Foundation.}} procedure DAOFIND is run on each $i$-band exposure to create master-list of sources, which is supplemented by $ugr$H$\alpha$ source detection's to capture any faint blue or H$\alpha$ bright sources. A point source function was fit using the DAOPHOT procedure to obtain photometry in all-bands in Vega magnitudes. AAVSO photometric all-sky survey{\footnote{https://www.aavso.org/apass}} (APASS) photometry is utilised to bring the survey photometry on a global calibration scale. To select sources with photometric accuracy we apply the following selection criteria-

\begin{enumerate}
\item $22$\,$>$\,$r$\,$>$\,13 in both the red and blue filter sets to avoid saturated and faint sources;
\item signal to noise ratio $>$\,10 in $ri$H$\alpha$ bands; 
\item point source function fit of $\chi$\,$<$\,1.5 to select stellar or star-like sources
\end{enumerate}

In the resulting sample, 2\,091\,573 unique sources have $ri$H$\alpha$ photometry meeting our quality criteria. The saturation limit of the VPHAS+ photometry ($r$\,$\sim$\,13\,mag) means that the upper main sequence of the cluster is saturated. 

\subsection{{\it Gaia} DR\,2 astrometry}
In addition to the VPHAS+ data, we make use of astrometric data from {\it Gaia}\,DR2 \citep{gaia}. This data contains parallaxes ($\pi$), and proper motions vectors in Right Ascension ($\mu_{\alpha}$) and Declination ($\mu_{\delta}$) calculated from the first two years of {\it Gaia} observations (2014-2016). The {\it Gaia} DR2 dataset has a detection threshold of $G\sim$\,20.5\,mag towards Sh\,2-012, although the very bright ($G<$\,7\,mag) and high proper-motion ($>$\,0.6 arcsec\,yr$^{-1}$) stars are incomplete. Global correlations on the parallax and proper motions are believed to be around $\pm$0.1\,mas and $\pm$0.1\,mas\,yr$^{−1}$. In addition, comparison with known quasars indicates an overall negative bias of 0.029\,mas \citep{gaia2} which we add to our parallax measurements. 
We cross-matched our VPHAS+ source list with the {\it Gaia}\,DR2 dataset within a radius of 0.1$\arcsec$ (the cross match radius was chosen experimenting with the astrometric fidelity of both catalogues). Finally, we applied the \cite{gaia2} C-1 astrometric equation to remove sources with unreliable astrometry. We impose the requirement of astrometric criteria on the VPHAS+ dataset as the cluster is in the direction of the Galactic plane from our viewpoint, and a significant proportion of background stars would be erroneously selected using purely photometry. In total, we have 1\,296\,410 stars with high-quality astrometry and photometry. These form the source dataset from which we will identify CTTS.

\subsection{Near and mid-Infrared imaging}

In addition to optical photometry, we make use of near and mid-infrared survey photometry to identify the disc stage of CTTS. To do, we cross matched our catalogue with the $ZYJHK$s VVV (Vista Variables in the Via Lactea; \citealt{vvv}) survey using a cross-match radius of 0.3$\arcsec$. We discarded sources having random errors $>$\,0.1\,mag in $JHK$s, and classified as having a non-stellar profile. In addition, we cross-matched our sample with the GLIMPSE (The Galactic Legacy Infrared Midplane Survey Extraordinaire) catalogue providing photometry at 3.6, 4.5, 5.8 and 8.0\,$\mu$m. The survey was conducted using the {\it Spitzer} space telescope. We cross-matched with our parent catalogue using a radius of 1$\arcsec$, and checking the $K$s magnitudes of stars given in the GLIMPSE survey catalogue (which are taken from the 2MASS survey of \cite{cutri03}) against the VVV provided $K$s magnitudes. In addition, we discarded stars having random photometric uncertainties in any band greater than 0.2\,mag.

\section{Classical T\,Tauri stars}

\subsection{Identifying CTTS using their H$\alpha$ photometric equivalent width} 

The ($r-i$) vs. ($r-$H$\alpha$) diagram (Fig.~\ref{fig:rha}) is used for identifying H$\alpha$ emission line stars. The ($r-$H$\alpha$) colour measures the strength of the H$\alpha$ line relative to the $r$-band photospheric continuum. It is only slightly affected by relative extinction, due to the negligible extinction coefficient between these two bands. For late-type stars, the ($r-i$) colour is a good proxy for spectral type, for stars with low extinction. Since most main-sequence stars do not have H$\alpha$ in emission, their ($r-$H$\alpha$) colour at each spectral type provides a template against which ($r-$H$\alpha$) colour excess caused by any H$\alpha$ emission can be measured. From \cite{Kalari15}, it is given by, ($r-$H$\alpha$)$_{\rmn{excess}}$\,=\,($r-$H$\alpha$)$_{\rmn{observed}}-$($r-$H$\alpha$)$_{\rmn{model}}$. The ($r-$H$\alpha$)$_{\rmn{excess}}$ is used to compute the H$\alpha$ equivalent width (EW$_{{\rmn{H}}\alpha}$) from the equation  

  \begin{equation}
    \ \rmn{EW}_{\rmn{H}\alpha}= \rmn{W}\times[1-10^{0.4\times(r-\rmn{H}\alpha)_{\rmn{excess}}}]\
  \end{equation}
following \cite{de10}. W is the rectangular bandwidth of the H$\alpha$ filter. The photometric EW$_{\rmn{H}\alpha}$ for all stars having $ri$H$\alpha$ photometry in our sample are thus measured.

We use the spectral type-EW$_{\rmn{H}\alpha}$ criteria of \cite{barr03} to select CTTS from the identified H$\alpha$ emission line stars. H$\alpha$ emission line stars having spectral type earlier than K5 and EW$_{\rmn{H}\alpha}$\,$<$\,$-18$\,\AA, K5-M2.5 and EW$_{\rmn{H}\alpha}$\,$<$\,$-25$\,\AA~ and M2.5-M6 and EW$_{\rmn{H}\alpha}$\,$<$\,$-38$\,\AA\ are selected. This criteria includes the consideration that the maximum errors due to a combination of the reddening uncertainty, and random photometric errors are around 9--12\,\AA\ depending on spectral type. The selection criteria is designed to exclude any possible interlopers such as chromospherically active late-type stars. We identify 156 CTTS on this basis. For a detailed comparison of the accuracy of photometric EW$_{\rmn{H}\alpha}$ to spectroscopically measured ones, the reader is referred to \cite{geert11} and \cite{Kalari15}. In addition, we present in electronic form the code and stellar models utilised for calculating photometric EW$_{\rmn{H}\alpha}$$^{*}$.

\begin{figure}
\center
\includegraphics[width=72mm, height=58mm]{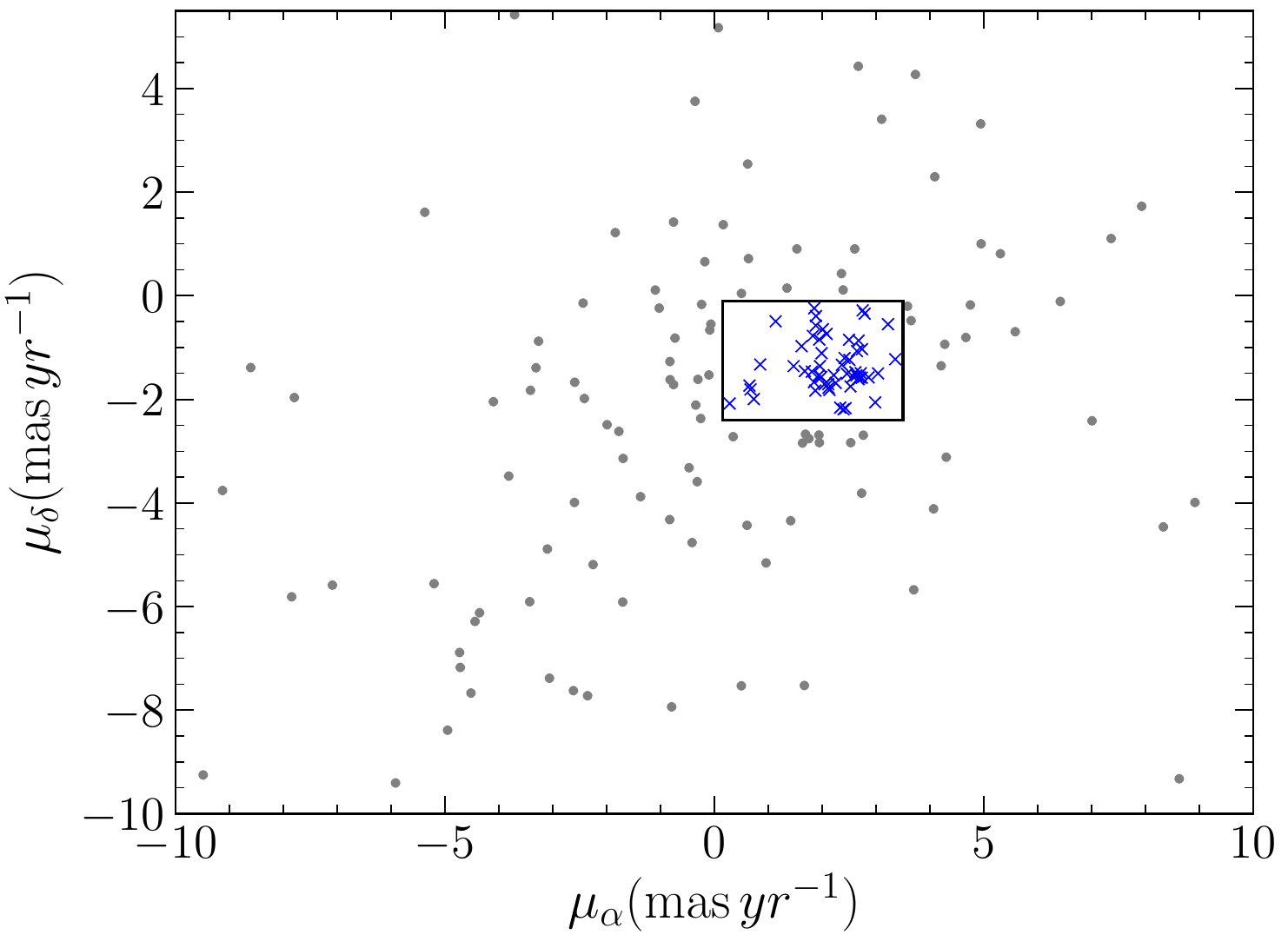}
\caption{Position of stars selected using the EW$_{\rmn{H}\alpha}$ criteria (grey circles) in proper motion ($\mu_{\alpha}$-$\mu_{\delta}$) space. The limits of the cluster are shown as a solid box, with the kinematically selected members shown as blue crosses.}
\label{kinematic}
\end{figure} 


\subsection{Removal of kinematic outliers}

A significant number of stars identified as CTTS based on their photometric EW$_{\rmn{H}\alpha}$ have kinematic properties statistically different from the mean of the population. These stars are very likely contaminant non-members that are found further in the spiral arm or located in front of the region \citep{Rauw10}, although a small fraction might be runaway or dynamically ejected stars that might be true members. However, we remain conservative in our choice and exclude all kinematic outliers as non-members. 

To select the kinematic outliers, we first identify the cluster population in proper-motion space. We fit to distributions of $\mu_\alpha$ and $\mu_\delta$ a double-peaked Gaussian, where the broad peak identifies the background population and the narrow peak the cluster population. The resulting fits give the centre of the cluster in proper motion space as $\mu_\alpha$=2.22$\pm$0.65\,mas\,yr$^{-1}$, $\mu_\delta$=$-$1.50$\pm$0.5\,mas\,yr$^{-1}$. In Fig.\,\ref{kinematic}, we show the selected CTTS in $\mu_\alpha$-$\mu_\delta$ space. The cluster exhibits a well defined centre, corresponding to the values found in the histograms. To define the cluster members, we select all stars within the inter-quartile range of the mean (this is resistant to outliers in the distribution), which are indicated by the solid lines. From this selection, 55 CTTS are selected as kinematic members of Sh\,2-012. On the sky, the kinematically selected sub-sample are clustered, where as the outliers are evenly distributed further suggesting that the applied rejection criteria is useful to remove background/foreground contaminants. Finally, we calculate the median distance of the kinematically selected population as $d$=1342$\pm$130\,pc based on {\it Gaia}\,DR2 parallaxes. This distance is in excellent agreement (within errors) with those measured based on analysis of the eclipsing binary V\,701\,Sco \citep{bell87}, and spectroscopic parallaxes \citep{Rauw08, Rauw10}. {\it Gaia}\,DR2 results rule out the high and low distances estimates of 1700 and 985\,pc from \cite{paun} and \cite{khar} respectively. The distance we adopt for our analysis is 1340\,pc.

\subsection{Stellar properties}

\begin{figure*}
\center
\includegraphics[width=90mm, height=75mm]{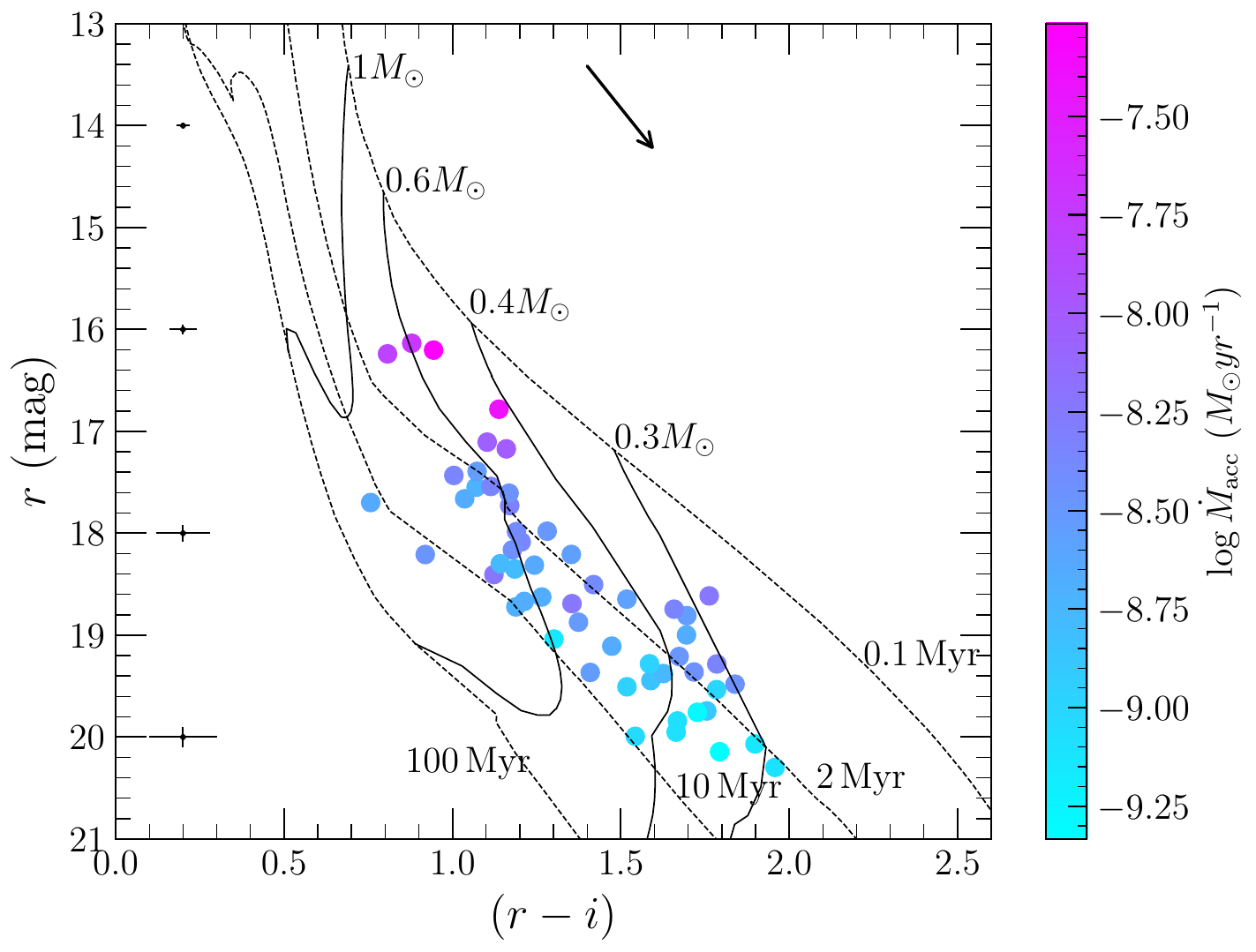}
\caption{$r$ versus ($r-i$) CMD of CTTS candidates. Labelled isochrones (dashed lines) and tracks (solid lines) are from Bressan et al. (2012). Model isochrones and tracks are reddened by $E(B-V)$\,=\,0.32 following a reddening law $R_{V}$\,=\,3.1. Circles indicate the position of each candidate CTTS, and colour is representative of log\,$\dot{M}_{\rmn{acc}}$ as indicated by the colorbar. The reddening vector for $A_{V}$\,=\,1 is shown. The median magnitude errors in a given 2\,mag bin, and the associated colour errors are shown at ($r-i$)\,=0.2 and the corresponding magnitude.}
\label{fig:cmd}
\end{figure*}

The stellar mass ($M_{*}$) and age ($t_{*}$) of each kinematically selected CTTS is estimated by interpolating its position in the observed $r$ versus $(r-i)$ CMD with respect to PMS tracks and isochrones (Fig.~\ref{fig:cmd}). We employ \cite{bress12} solar metallicity single star isochrones and tracks, and compare the estimated parameters with those estimated using the \cite{siess00} and \cite{pisa} stellar models. We assume the mean reddening $E(B-V)$\,=\,0.32, and a standard Galactic reddening law of $R_{V}$\,=\,3.1, although this may not be reflective of the entire cluster \citep{Rauw10}, but only of the bulk population. The \cite{pisa} and \cite{siess00} models were converted from the theoretical plane to the VPHAS+ magnitudes using the ATLAS\,9 stellar models \citep{cast04}. Errors on the interpolated values were calculated by propagating the photometric uncertainties, and including an absolute extinction and distance uncertainty of 0.2\,mag and 100\,pc respectively. 
 
\begin{figure} 
\center
\includegraphics[width=80mm, height=135mm]{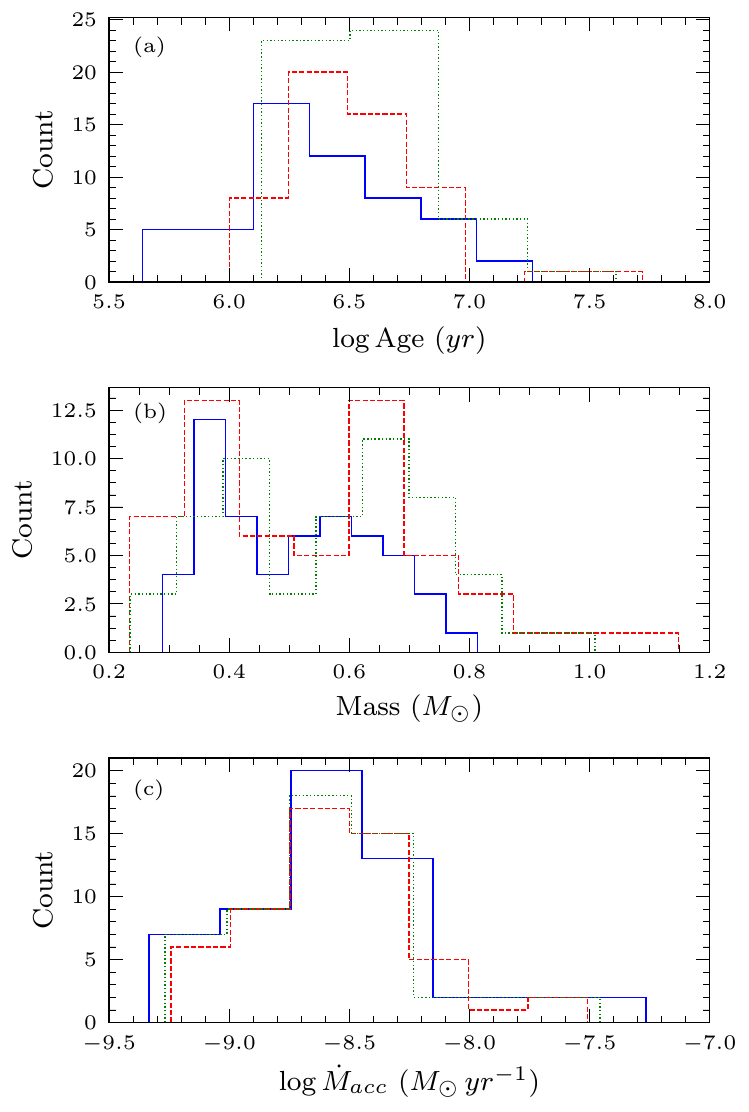}
\caption{The distribution of masses (a), ages (b) and $\dot{M}_{\rmn{acc}}$ (c) determined using different stellar evolutionary models. Solid blue lines are results estimated using the Bressan et al. (2012) models, dashed red lines using the Tognelli et al. (2011) models and dotted green lines using the Siess et al. (2000) models.}
\label{fig:comphist}
\end{figure}

The mean ages of CTTS derived from the \cite{bress12} isochrones is 2.8\,$\pm$\,1.6\,Myr, and the range is between 0.4\,-\,18\,Myr (Fig.~\ref{fig:comphist}a), with an interquartile range between 1.4--3.8\,Myr.
The error on the mean age takes into account the reddening uncertainties and photometric errors. The standard deviation of CTTS ages is 3.4\,Myr, and 70\,\% of all CTTS have ages between 1\,--4\,Myr. The individual ages of CTTS vary with the isochrones used (see Fig.~\ref{fig:comphist}a). The mean age estimated using the \cite{pisa} (2.9\,$\pm$\,1.7\,Myr) and the \cite{siess00} (3.5\,$\pm$\,2.4\,Myr) isochrones is higher than the \cite{bress12} isochrones. It is important to stress that the statistical age of a CTTS population is considered relatively accurate compared to the individual ages of CTTS due to the combination of reddening and distance uncertainties \citep{mayne08}. The differences in ages derived from different isochrones follows the pattern observed in \cite{herc15}. The authors of that paper found that for stars with spectral types earlier than M (the bulk of our sample), the \cite{bress12} isochrones give a smaller age than the \cite{siess00} isochrones. 

Our results agree well with literature age estimates of stars in the central open cluster NGC\,6383. In particular, the age of NGC\,6383 X-ray emitting stars derived by \cite{Rauw10} is 2.8$\pm$0.5\,Myr. The ages of cluster members identified using photometric criteria by \cite{paun} are between 1\,--\,4\,Myr, although we caution that their distance estimate is significantly higher (1700\,pc). \cite{fitz} quote an age of 1.7$\pm$0.4\,Myr for bright cluster members, assuming a distance of 1500\,pc. The central O7\,V binary HD\,159176 has an age of 2.8\,Myr according to \cite{fitz} and around 2.3\,--\,2.8\,Myr following \cite{Rauw10}, which all agree well with our mean ages. It is important here to note that both \cite{Rauw10} and \cite{paun} ages are derived for photometrically identified members. In the case of \cite{Rauw10}, the authors derive the age using X-ray identified cluster members. It is also generally accepted that X-ray emitting PMS stars (likely Weak Line T\,Tauri stars) are generally older than CTTS.


The interpolated mass distribution is shown in Fig.~\ref{fig:comphist}b. The masses determined using different isochrones agree well within the errors for individual stars (barring few outliers). The mass range is between 0.3--0.9\,$M_{\odot}$, with the median mass 0.5\,$M_{\odot}$. The lower mass cut-off corresponds to the H$\alpha$-band cutoff, and we are relatively complete at masses above 0.4\,$M_{\odot}$ for the 2\,Myr isochrone.      

\subsection{Accretion properties}

It is generally accepted that the stellar magnetosphere truncates the circumstellar disc at an inner radius ($R_{\rmn{in}}$). Matter is transferred along the magnetic field lines at this truncation radius, and releases energy when impacting on the star. The energy generated by magnetospheric accretion ($L_{\rmn{acc}}$) heats and ionises the circumstellar gas, thereby causing line emission. The reradiated energy that has gone towards ionising the gas can be measured from the luminosity of the line emission, which is correlated to $L_{\rmn{acc}}$ (\citealt{Hart16}). Assuming matter is free-falling from $R_{\rmn{in}}$, the $\dot{M}_{\rmn{acc}}$ can be estimated using the free-fall equation from $R_{\rmn{in}}$ to the $R_{\ast}$ if one knows the $M_{\ast}$ and energy released, i.e. $L_{\rmn{acc}}$. We can therefore measure the H$\alpha$ line luminosity (${L_{\rmn{H}\alpha}}$) from the EW$_{{\rmn{H}}\alpha}$ to estimate the $L_{\rmn{acc}}$ and $\dot{M}_{\rmn{acc}}$. 

To estimate the ${L_{\rmn{H}\alpha}}$, we first estimate the H$\alpha$ line flux ($F_{\rmn{H}\alpha}$) by subtracting the H$\alpha$ continuum flux ($F_{\rmn{continuum}}$) from the total flux given by the H$\alpha$ unreddened magnitude. We refer the reader to \cite{Kalari15} for details on how the $F_{\rmn{H}\alpha}$ is calculated. Following from \cite{de10}, we assume 2.4 per\,cent of the H$\alpha$ intensity is caused by contamination from the N[{\scriptsize II}] $\lambda$$\lambda$ 6548, 6584\,\AA\,lines and subtract it by that amount. 

$L_{\rmn{acc}}$ is related to $L_{\rmn{H}\alpha}$ as:
\begin{equation} 
\rmn{log}~L_{\rmn{acc}} = (1.13\pm 0.07)\rmn{log}~L_{\rmn{H}\alpha}+(1.93\pm 0.23)
\end{equation}
\citep{geert11}. Here, $\rmn{log}~L_{\rmn{acc}}$ and $\rmn{log}~L_{\rmn{H}\alpha}$ are measured in units of solar luminosity, $L_{\odot}$. The root mean scatter is 0.54\,dex. The scatter in the observed relationship is likely caused by a combination of circumstellar absorption, variability in accretion, and unrelated sources of line emission. 

The $\dot{M}_{\rmn{acc}}$ can then be estimated from the free-fall equation: 
  \begin{equation}
    \dot{M}_{\rmn{acc}}= \frac{L_{\rmn{acc}}R_{\ast}}{\rmn{G}M_{\ast}}\Bigg(\frac{R_{\rmn{in}}}{R_{\rmn{in}}-R_{\ast}}\Bigg).
  \end{equation} 
$M_{\ast}$ and $R_{\ast}$ are the stellar mass and radius respectively. $R_{\rmn{in}}$ is the truncation radius. We adopt $R_{\rmn{in}}$\,=\,5\,\,$\pm$\,\,2\,$R_{\ast}$ 
from (\citealt{gullbring98, vink05}). The resultant distribution of mass accretion rates are plotted in Fig.~\ref{fig:comphist}c. The median logarithm of $\dot{M}_{\rmn{acc}}$ is $-$8.5\,$M_{\odot}yr^{-1}$. In addition to the propagated uncertainties in the interpolated stellar mass and radii, the major error on the $\dot{M}_{\rmn{acc}}$ is due to the scatter in the $\rmn{log}~L_{\rmn{acc}}$--$\rmn{log}~L_{\rmn{H}\alpha}$ relation. The estimated results are given in Table 1.

\subsection{$u$-band accretion properties}

As discussed in Section 3.4, it is thought that mass infall along the magnetic field lines results in an accretion shock as matter hits the stellar surface. This phenomena is seen not only in the reradiated energy caused from the ionisation lines, but also in the excess continuum luminosity at short wavelengths due to the accretion shocks. Similarly to the method described in Section 3.4, the $L_{\rmn{acc}}$ is correlated to the excess luminosity but at continuum ultraviolet wavelengths. 

\begin{figure*}
\center
\includegraphics[width=175mm, height=65mm]{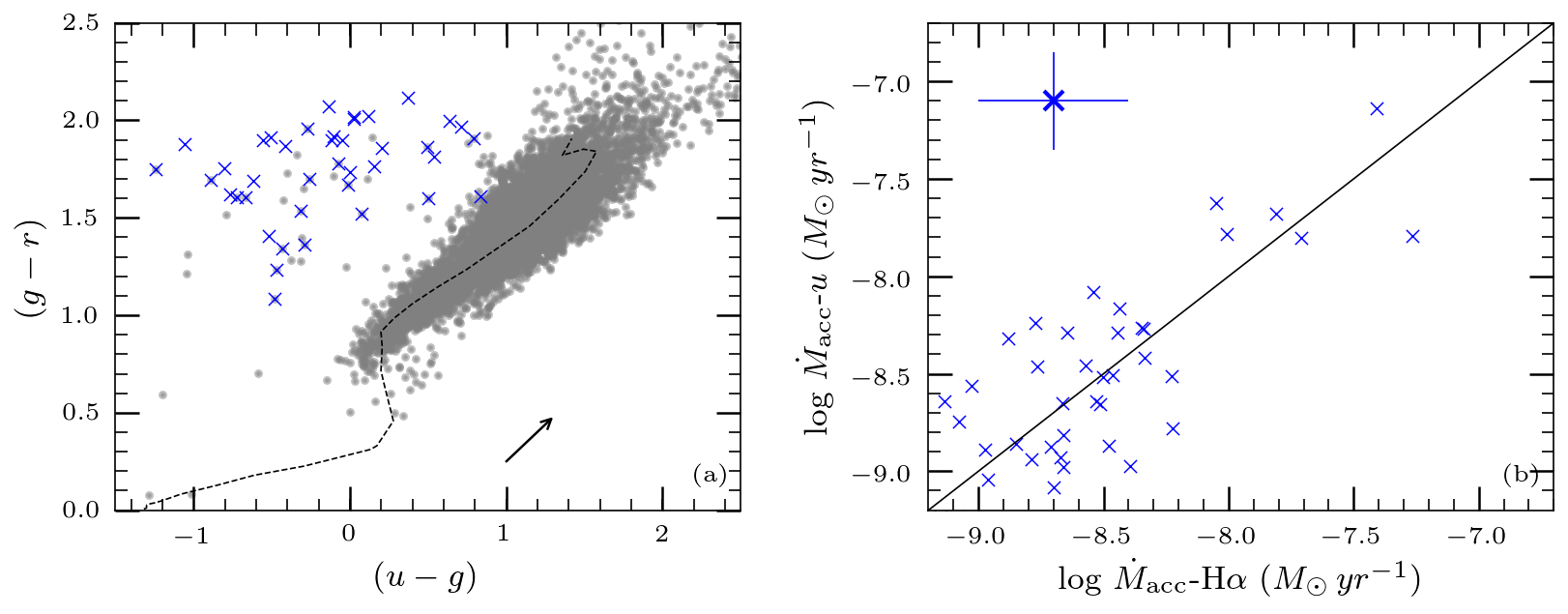}
\caption{(a) ($u-g$) versus ($g-r$) diagram. The dashed line is the reddened model colour track from Castelli \& Kurucz (2004). Stars meeting our selection criteria (grey dots) and CTTS identified on the basis of their EW$_{\rmn{H}\alpha}$ (blue crosses) are shown. A reddening vector for $A_{V}$\,=\,1 is shown. (b) Comparison of $\dot{M}_{\rmn{acc}}$ derived from H$\alpha$ line luminosity and $u$-band excess luminosity. Mean error bars are shown in the upper left hand corner.}
\label{fig:ugr} 
\end{figure*}

Forty CTTS have $ugr$ photometry (i.e. the blue filter set described in Section 2). Their positions in the ($u-g$) vs. ($g-r$) diagram are shown in Fig.~\ref{fig:ugr}a. To measure their excess $u$-band emission, we take the $(g-r)$ colour as a proxy for spectral type. The excess emission flux ($F_{u,\rmn{excess}}$) is given by  
  \begin{equation}
    \ F_{u,\rmn{excess}}= F_{0,u}\times[10^{-u_{0}/2.5}-10^{-(u-g)_{\rmn{model}}+g_{0})/2.5}].\
  \end{equation}
$F_{0,u}$ is the $u$-band integrated reference flux. $u_{0}$ and $g_{0}$ are the dereddened magnitudes, and $(u-g)_{model}$ is the corresponding model colour from \cite{cast04}. The excess flux is used to estimate the $L_{\rmn{acc}}$ using the empirical relation \citep{gullbring98}
\begin{equation} 
\rmn{log}~L_{\rmn{acc}} = {log}~L_{u,\rmn{excess}}+0.98.
\end{equation}
$L_{u,\rmn{excess}}$ is the $u$-band excess luminosity. $L_{\rmn{acc}}$ and $L_{u,\rmn{excess}}$ are in units of solar luminosity $L_{\odot}$. Mass accretion rates are calculated using Eq.~2, with results compared to H$\alpha$ derived $\dot{M}_{\rmn{acc}}$ in Fig.\,\ref{fig:ugr}b.

On comparison, we find that the mean scatter from the 1:1 relation between H$\alpha$ and $u$-band derived accretion rates is 0.3\,dex. Considering that the accretion variability is around 0.5\,dex \citep{cost12}, the scatter observed maybe due to a combination of variability (since the optical and ultraviolet observations are non-contemporaneous; see Section 2) and contamination in the measured accretion luminosity. These are however within the scatter measured spectroscopically using a variety of contemporaneously measured line and continuum indicators \citep{manara17}. This is approximately similar to the result found in \cite{Kalari15}. It indicates that the use of accretion luminosities derived from H$\alpha$ photometry is comparable to oft-used line luminosities from spectroscopy, or $u$-band photometry, and is a reliable indicator of accretion line intensity. The availability of H$\alpha$ photometry with large surveys such as VPHAS+ now allows to determine efficiently accretion rates across the plane of the Milky Way. 

\subsection{Disc properties}
\subsubsection{Near-infrared properties}
   
CTTS display near-infrared excesses when compared to a purely stellar template due to the presence of dust in their inner circumstellar discs \citep{cohen79}. This allows for a simple independent check on our H$\alpha$ identified CTTS.

We identify near-infrared {\it{JHK}}s counterparts of the CTTS in the VVV survey as described in Section 2.3. The limiting magnitude of the VVV survey depends on the region of sky observed. Around Sh\,2-012, VVV photometry reaches a depth of $K$\,$\sim$\,18\,mag, which translates to roughly a star of spectral type K7, suggesting that a considerable number of the stars in our sample should be identifiable in the VVV survey. 51 stars in our sample of 55 were found to have VVV counterparts with high-quality photometry. 

In Fig.~\ref{fig:ir}a, the ($J-H$) vs. ($H-K$) colour-colour diagram of CTTS in Sh\,2-012 are plotted. The dashed line is the CTTS locus of \cite{meyer97} which predicts the excess emission using disc accretion models having log\,$\dot{M}_{\rmn{acc}}$ between $-$6 to $-8$\,$M_{\odot}yr^{-1}$. The solid line is the main sequence locus from \cite{bess98}. Most stars ($\sim$\,94\,per\,cent) having near-infrared colours in our sample have ($H-K$s) excesses similar to the CTTS locus, suggesting on the basis of the near-infrared diagram alone that they are CTTS. This confirms our sample as consisting primarily of CTTS undergoing accretion. Five stars lie near the main sequence locus, indicating that they are either weak-line T\,Tauri stars, or have no excess infrared emission suggestive of a circumstellar disc, and might be interlopers. 


\subsubsection{Mid-infrared properties}

\begin{figure*}
\center
\includegraphics[width=175mm, height=65mm]{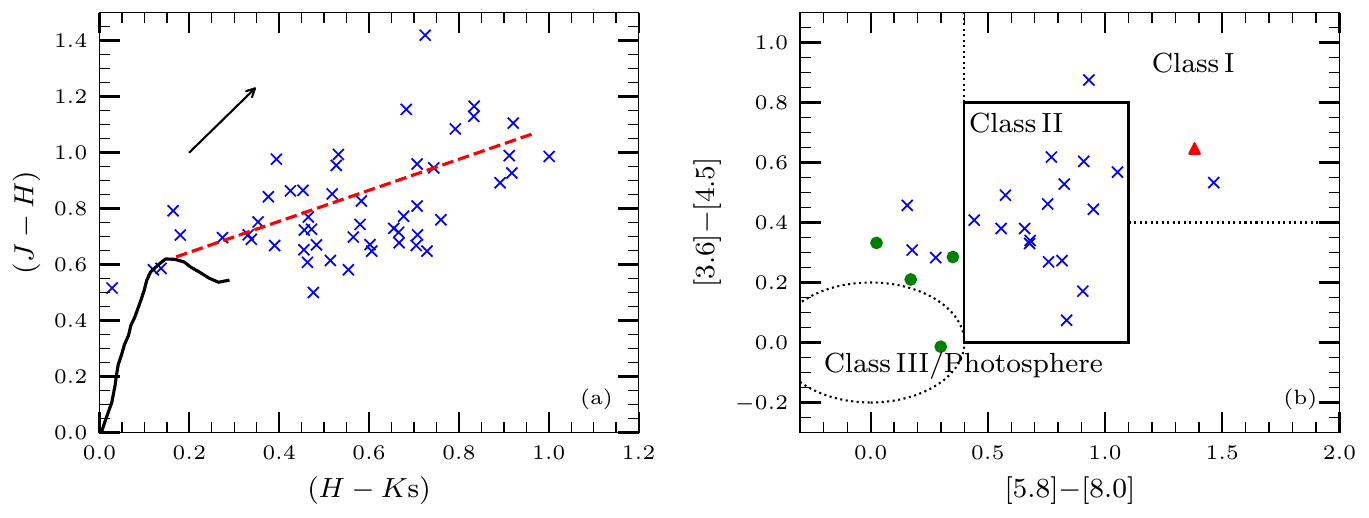}
\caption{The CTTS candidates in the near-infrared (a) and mid-infrared (b) colour-colour diagrams are shown. In (a), the solid line is the locus of main-sequence stars from Bessell et al. (1998), and the dashed line the CTTS locus of Meyer et al. (1997). The reddening vector of $A_V$=1 is also shown. (b) shows the 3.6\,$\micron-$4.5\,$\mu$m vs. 5.8\,$\mu$m$-$8\,$\micron$ colour-colour diagram, with solid line demarcating the colours of Class II CTTS having accretion rates of $10^{-8}$\,$M_{\odot}yr^{-1}$. The regions corresponding to Class I and Class III sources are also marked and labelled.}
\label{fig:ir}
\end{figure*}

We also found mid-infrared counterparts in the {\it{Spitzer}} IRAC data described in Section 2.2. 26 cross-matches within 1$\arcsec$ having high-quality photometry were found. In Fig.~\ref{fig:ir}b, the 3.6\,$\micron-$4.5\,$\mu$m vs. 5.8\,$\mu$m$-$8\,$\micron$ colour-colour diagram is plotted. The solid box represents the typical colours for CTTS having mean $\dot{M}_{\rmn{acc}}$\,=\,$10^{-8}$\,$M_{\odot}yr^{-1}$, while the areas occupied by the Class\,I and Class\,III sources are also shown. Most of the cross-matched CTTS candidates in our sample lie in this box, providing an additional sanity check on the results. 

\begin{figure}
\center
\includegraphics[width=78mm, height=60mm]{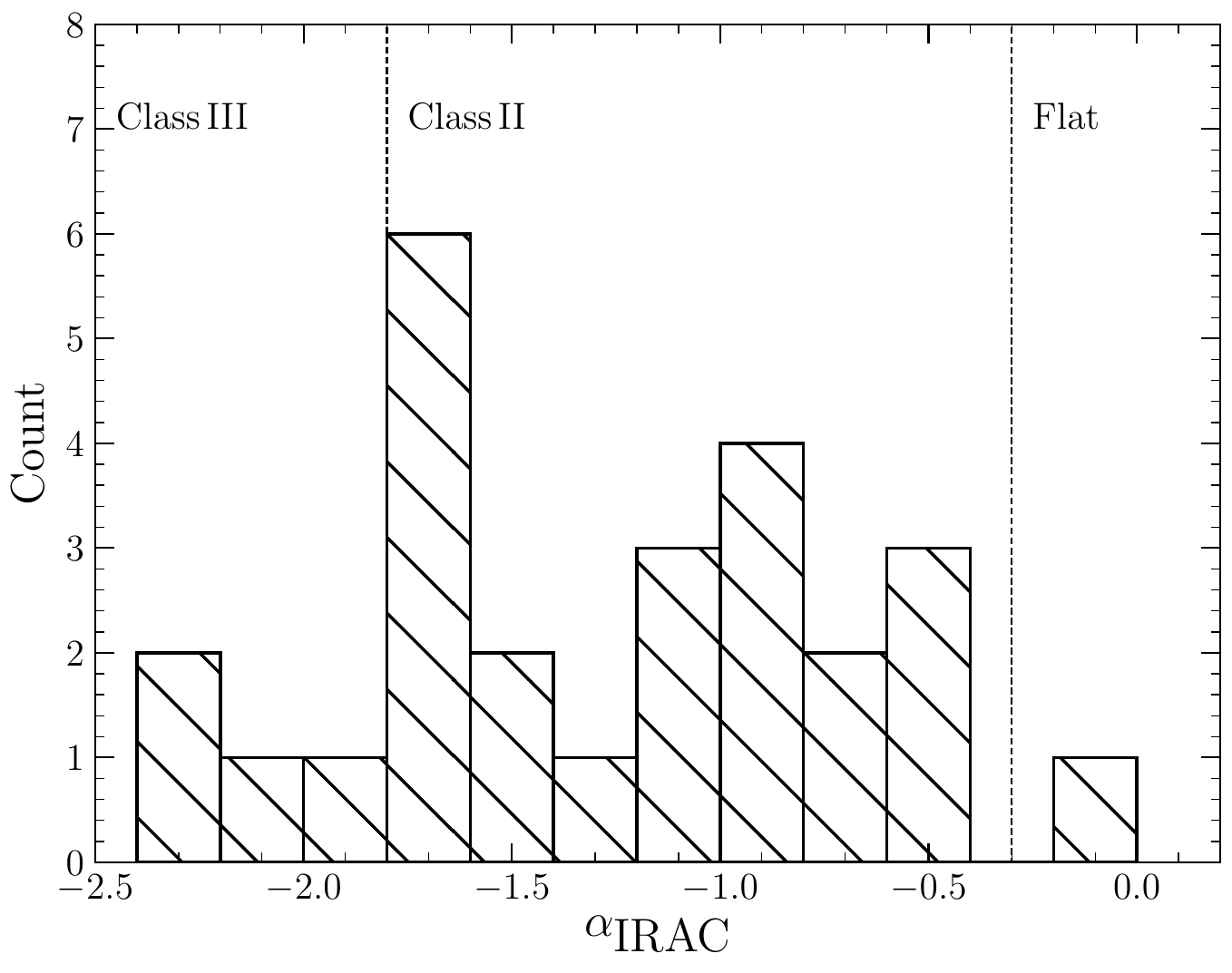}
\caption{Histogram of SED slopes of CTTS. Only 26 CTTS candidates having cross matches in {\it{Spitzer}} photometry are shown, with 21 sources identified as Class II sources, 4 as Class III sources and 1 as a Flat spectrum source.}
\label{slope}
\end{figure}

The slope of the spectral energy distribution (SED) in the infrared,
\begin{equation} 
\alpha_{\rm{IR}}\,=\,\frac{d\,{\rm{log}}(\lambda F_{\lambda})}{d\,{\rm{log}}\lambda},      
\end{equation}
(at $\lambda$\,$>$\,3\,$\micron$) is used to diagnose the evolutionary
stage of the disk-star system \citep{lada76}, and provides another sanity check on the presence of a circumstellar disc around the CTTS. We adopt the classification
scheme of \cite{greene} to distinguish between systems with
protostellar disks (Class I), optically thick disks (Class II), transitional/anaemic disc (Class III), or main-sequence stars. $\alpha_{\rm{IR}}$ was derived by
fitting the {\it{Spitzer}} magnitudes at 3.6, 4.5, 5.8, and
8$\micron$. The resultant $\alpha_{\rm{IR}}$ values are given in Fig.\,\ref{slope}.

Out of the 26 accreting PMS stars that have {\it{Spitzer}} photometry at all
observed wavelengths, we find that 1 has a SED slope resembling a Flat spectrum source, and 21 have SED slopes resembling Class\,II YSOs. The remaining 4 stars have SED slopes of Class\,III objects, resembling a transitional/anaemic disc. Overall, most of the PMS
star sample have infrared-excess akin to circumstellar disks, verifying
that we deal with genuine accreting PMS stars. Overall, the results from the IRAC colour-colour diagrams correlate well with the SED slopes. 

\subsection{Contaminants}

From infrared photometry, we find that five of our stars do not have infrared colours resembling CTTS. These five stars fall on the main sequence locus in the ($J-H$) vs. ($H-K$s) colour-colour diagram. On inspection, they have distances varied more than 3$\sigma$ from the mean according to {\it Gaia} data. Their proper motions, or spatial locations are not clustered, and they are distributed near the central locus of stars in both diagrams. Neither are their locations in the colour-magnitude diagram deviant from the main bulk of stars.

Neglecting that this difference is not due to incorrect cross-matches, or accretion variability, we suggest that one in approximately 10 stars is a likely contaminant in our sample. This would put the total contaminants in our sample to be around $\sim$10\%, or 6. Our contamination is expected to be very low, as we have used exclusive EW$_{\rm{H\alpha}}$ criteria to remove chromospheric outliers. In addition, the utilisation of proper motions from {\it Gaia} DR2 was essential in removing the bulk of the foreground/background population observed in NGC\,6383 by \cite{Rauw10}. This ensures that we have a relatively clean sample of CTTS in Sh\,2-012 for which we have determined the stellar and acccretion properties.

\section{Discussion}

\subsection{$\dot{M}_{\rmn{acc}}$ and stellar properties}

\begin{figure}
\center
\includegraphics[width=78mm, height=60mm]{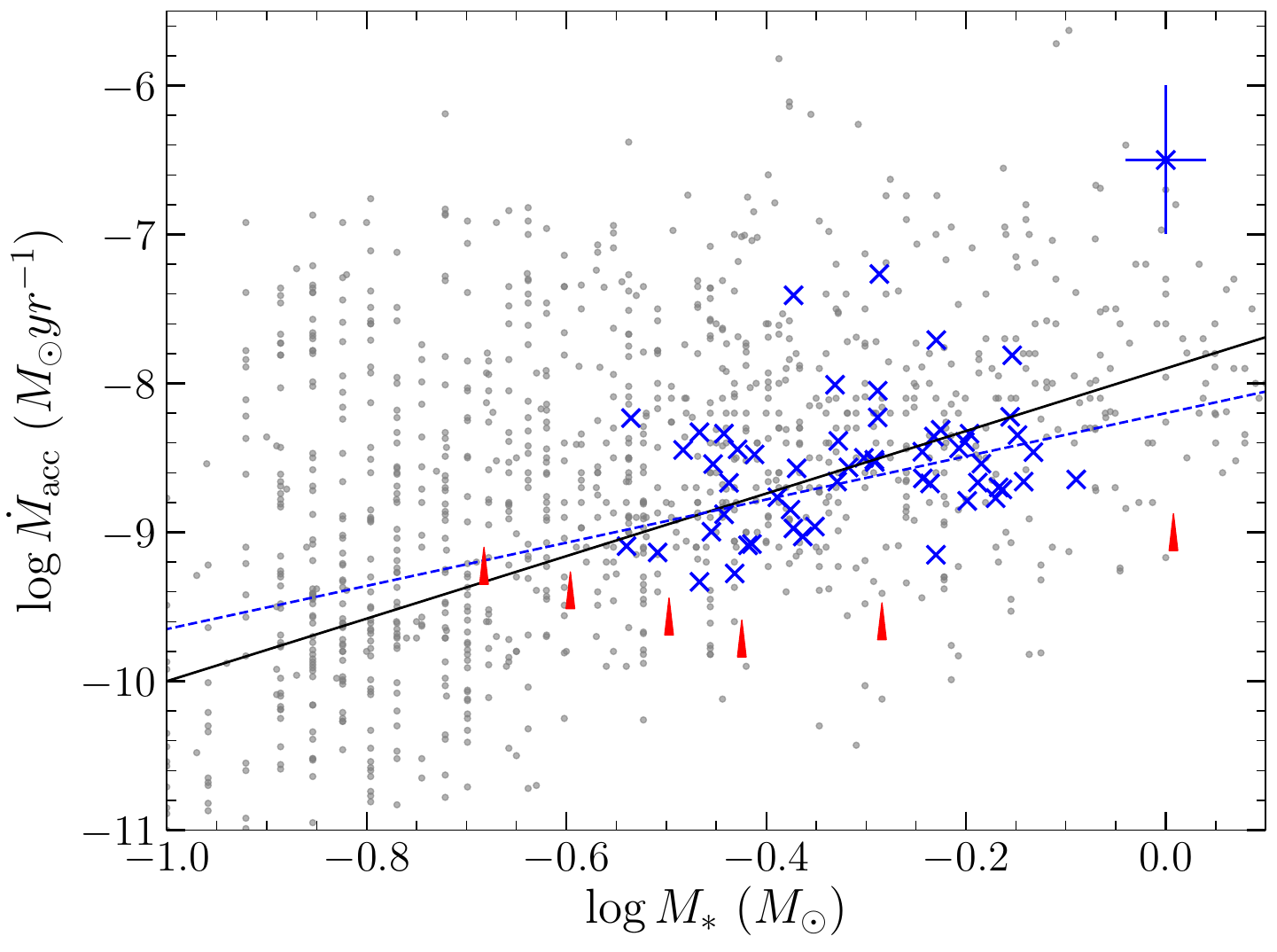}
\caption{$\dot{M}_{\rmn{acc}}$-$M_{\ast}$ relation of our sample. CTTS in Sh\,2-012 are shown as blue crosses, with red arrows marking the detectability limits. The blue dashed line is the best-fit relation for only the Sh\,2-012 stars, following a slope of 1.45. The fit to the literature data shown as solid circles has a slope of 2.1 (solid line). The difference in the intercept of the fit is 0.27\,dex}
\label{maccm}
\end{figure}

The $\dot{M}_{\rmn{acc}}$ of CTTS has been observed to correlate with the stellar mass, following a power-law form of $\dot{M}_{\rmn{acc}}$\,$\propto$\,$M_{\ast}^{\alpha}$. Although a double power-law has been recently suggested with the break at masses$\sim$ 0.3\,$M_{\odot}$ \citep{manara17}. From the literature, $\alpha$ is known to vary between 1--3 \citep{Hart16} in the mass range of 0.1--2\,$M_{\odot}$. From a survey of the literature results, the best fit across this mass range is found to be $2.1$, and the resultant fit is plotted in Fig.\,\ref{maccm} (see \citealt{thesis} for details of the literature results). Overplotted are the accretion rates of CTTS in Sh\,2-012. The best-fit relation is 1.45$\pm$0.3, which was calculated using a linear-regression fit following the Buckley-James method accounting for the limits of our detection, and errors on both variables. The detection limits on the data were calculated assuming the lower limits of our EW$_{\rm{H\alpha}}$ criteria, and a 2\,Myr isochrone.

For the stars in Sh\,2-012, the range of $\dot{M}_{\rmn{acc}}$ falls well within those observed in the literature at any given mass. Based on this, and the detection limits of our study, it is apparent that we are unable to detect small accretion rates at the lowest masses, which would explain the shallowness of our slope compared to the literature. In addition to the observed slope, there is a scatter of around 1\,dex at any given mass in our sample. Much of the observed scatter in this relation is real as the amount that can be attributed to variability is too small ($\sim$0.5\,dex; \citealt{cost12}). Additionally, the variation in the slope cannot be attributed in our case to differences between the stellar models. At these masses (0.3--1\,$M_{\odot}$), differences between stellar models are smaller than at lower and higher masses \citep{herc15}, resulting in a negligible difference in the estimated stellar masses and accretion rates (Fig.\,\ref{fig:comphist}). The resulting value of $\alpha$ using the \cite{siess00} and \cite{pisa} models are 1.5 and 1.41 respectively. The differences between the slopes using different stellar models are much smaller than the error on the resultant fits.


Interestingly, the observed relation may indicate favoured methods of disc dispersal. In particular, the determined slope in the $\dot{M}_{\rmn{acc}}$--$M_{\ast}^{\alpha}$ relation. The value of $\alpha$ from our study is is different from the simple viscous disc evolutionary models of \cite{hart06}. It is consistent with viscous models combined with X-ray induced ionisation \citep{ercolano},which suggest a value of 1.7. But, as seen earlier our determined slope maybe affected by detection limits. Additionally, our values fall entirely within the range of previously determined $\dot{M}_{\rmn{acc}}$ from the literature, which also indicates that the measured slope could be due to detection limits. It is also the reason we do not attempt to fit a broken power law suggested by \cite{manara17} (additionally, the break happens at the lower end of stellar mass limit). 

However, the difference in the intercept of the CTTS in Sh\,2-012 compared to younger 1--2\,Myr old CTTS from the literature is around 0.3\,dex, which is entirely consistent with the the viscous accretion model \citep{hart06}. In the viscious disc accretion paradigm, differences in the overall age of the region with respect to much younger regions ($<$2\,Myr CTTS) would result in the lower number of accretors, and smaller accretion rates in Sh\,2-012. However, the large spread, and difference in ages between various stellar models prevents us from exhaustively testing this scenario. 

An interesting explanation of the difference in intercept is the idea that mass is gained by the circumstellar discs from the environment after initial collapse. The disks in this scenario transport material from the environment and onto the star, acting as a conveyor belt for interstellar material of the molecular cloud \citep{manara18}.  In this scenario, both the accretion of matter onto the star, and matter from the environment on the disc follow Bondi-Hoyle accretion where $\dot{M}_{\rmn{acc}}$--$M_{\ast}^{2}$ \citep{throop}. However, the intercept of the slope may be higher or lower depending on the environment, and the mean of $\dot{M}_{\rmn{acc}}$ is higher than the median. Given that region is relatively older (at roughly 3\,Myr) than most nearby star-forming regions around 1\,Myr \citep{Hart08}, we can surmise that possibly the lower accretion rates if not due to the higher age of the region might be due to smaller amount of surrounding material from which discs gain mass to accrete. If we assume that this value is an average depending on global molecular cloud density within a certain region rather than dependent on local over densities, the intercept of the $\dot{M}_{\rmn{acc}}$--\,$M_{\ast}$ relation might vary depending on this. We stress therefore for the need of sub-mm imaging of regions alongside the growing large dataset of optical-infrared photometry and astrometry to measure disc and molecular cloud properties associated to CTTS, to understand how discs accrete and evolve.

\begin{figure*} 
\center
\includegraphics[width=170mm, height=140mm]{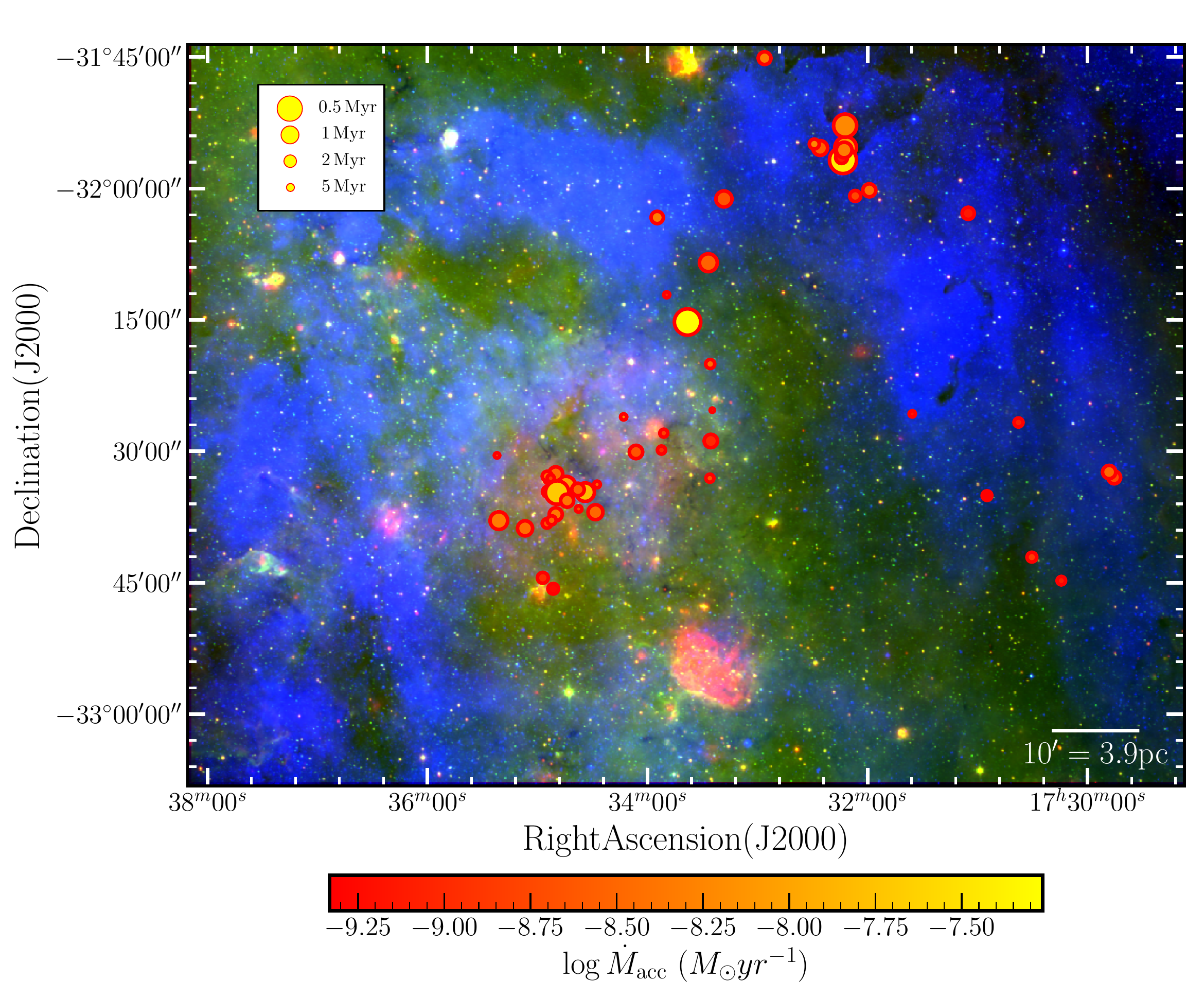}
\caption{Spatial location of CTTS in Sh\,2-012, with size of the symbols corresponding to the age according to the lower left legend, and the colour to the $\dot{M}_{\rmn{acc}}$ following the colour bar at the bottom. The positions are overlaid on a combination of H$\alpha$ (blue) showcasing the nebular and ionised structure of the region, {\it Spitzer} 8$\mu$m emission highlight the PAHs ionised from UV radiation (green), and a {\it Spitzer} 24$\mu$m image presenting the dust distribution in the region (red). Overall, the CTTS are located at the edges of the ionised/nebular structure.}
\label{maccplt}
\end{figure*}

\subsection{Spatial distribution of CTTS candidates}

Sh\,2-012 is a star-forming region centered on the open cluster NGC\,6383. The region has an approximate diameter of 35\,pc, with rims bright in H$\alpha$ visible towards the edges. Multiple dark cavities in the optical are noted towards the east and north of the cluster, bounded by the brightly ionised rims. We plot the distribution of CTTS on the sky overlaid with a three-colour image highlighting the morphology in H$\alpha$, PolyAromatic Hydrocardon (PAH) emission (8$\mu$m), and dust (24$\mu$m) in Fig.\,\ref{maccplt}. We find that the CTTS can be isolated into three different sub-regions within the larger region. Firstly, the central cluster contains the bulk of the CTTS ($\sim$42\%). Near the northern-western edge, towards a ionised bright rim at $\alpha$ of 17$^h$32$^m$16$^s$44 and $\delta$ of $-31^d$55$'08''$ there exists the other significant clustered population of CTTS ($\sim$22\%). It is noted that although these regions are separated by $\sim$17\,pc, the CTTS within them exhibit similar ages ($\sim$2\,Myr), and have the highest measured accretion rates. Finally, a spread of stars towards the western edge of the nebula, and in the interface between the northern-western edge and the central cluster contain the remaining CTTS. Interestingly, we note regions devoid of CTTS corresponding to dark cavities in H$\alpha$, but not with excess dust emission highlighting that the lack of CTTS is not likely due to obscuring dust. Overall, we detect two significant clustered population of CTTS around the central cluster and the north-west edge. They exhibit densities of around $\sim$5 CTTS\,pc$^{-1}$. The remaining stars are spread out towards the region west of the central cluster in the north-south direction. Comparing the position of the CTTS, and the morphology of the region a shell-like expansion structure is noted, with a bright rim at the extreme edge containing a significant population of CTTS. Such rims have been suggested as the location for future star-formation, likely triggered by the central cluster \citep{Rauw08}.

An useful way to test such scenarios is by comparing the spatial distribution of accretion rates. This circumvents the uncertainties in individual ages of stars. Underpinning this analysis is the assumption that the distribution of accretion rates is due to
the star formation history within a particular region. Younger
CTTS with higher accretion rates are located closer to their birth locations, while older CTTS may have had time to dynamically evolve and move away from their natal molecular cloud leading to larger spacings, and lower accretion rates. Assuming the $\dot{M}_{\rmn{acc}}$ differs by around $\sim$0.3\,dex in our mass range (Fig.\,\ref{maccm}), larger variations are reflective of dependencies on stellar age. Therefore, the distribution of $\dot{M}_{\rmn{acc}}$ maybe reflective of intrinsic age spreads.

Fig.\,\ref{maccplt} shows the distribution of CTTS in the sky with the $\dot{M}_{\rmn{acc}}$ of individual stars indicated by the hue of each symbol. From this, we find that the strongest accretors are concentrated around the central open cluster. In addition, the strongest accretors are also found towards the north-western edge, around a bright rim. If we consider the simple scenario that the central cluster formed first, triggering the formation of stars in the edges, there would be noticeable age difference between the two populations. Considering the upper limit on the age of the most massive central O7\,V star (2.3--2.8\,Myr; \citealt{Rauw10}), and the ages of the accretors in the bright rims there is a difference of around $\sim$1\,Myr. The distance between these regions is $\sim$17\,pc, suggesting that a cloud expansion speed of 17\,km\,s$^{-1}$ is necessary to trigger the observed population of CTTS, which is higher than average sound speeds in H{\scriptsize II}. Based on the fact that the accretion rates and ages of CTTS in the rims, and the central cluster are similar, we suggest that the morphology observed in Sh\,2-012 is not a result of triggered or sequential star formation, and that the stars across the whole region appear to have similar ages and are likely to have formed in a single burst of star formation around 2-3\,Myr ago. 

\section{Summary}

Based on H$\alpha$ excess emission, we identified 55 CTTS in the star-forming region Sh\,2-012. The identified CTTS have an age of 2.8\,Myr, and fall between 0.3--0.9\,$M_{\odot}$. From their H$\alpha$ and $u$-band excess intensities, we measured their mass accretion rates. The accretion rates correlate well with a scatter of 0.3\,dex, indicating the accuracy of $\dot{M}_{\rmn{acc}}$ measured using H$\alpha$ photometry which is now accessible for the entire Galactic plane with VPHAS+ survey data. The identified CTTS correspond well with the location of circumstellar disc bearing stars in the near and mid-infrared colour-colour diagrams, with the infrared SED slopes of all stars (having mid-infrared photometry) indicative of a circumstellar disc. 

When plotting the distribution of $\dot{M}_{\rmn{acc}}$ against $M_{\ast}$, we find a lower slope (although this also partly in cause due to our detection limits) and intercept than compared with the literature, and propose multiple explanations for this result, in line with known protoplanetary disc evolution theories. Viscous disc accretion may explain the smaller intercept, while X-ray photoevaporation can explain the observed slope in the relation. Finally, it is intriguing that the observed result may also be because protoplanetary discs accrete mass from the environment after collapse. The combination of VPHAS+, current infrared sky surveys with measurements of the dust distribution at sub-mm wavelengths in multiple-star forming regions can truly identify if such differences exist. 

Finally, examining the distribution of CTTS on the sky, we find that CTTS are concentrated in the central cluster, and towards the bright rims. The stars have similar ages, and accretion rates which favour a scenario of a single burst of star formation.

\section*{Acknowledgements}
We thank the anonymous referee, J.E. Drew, and J.S. Vink for constructive comments. V. M. K. acknowledges funding from CONICYT Programa de Astronomia Fondo Gemini-Conicyt as a GEMINI-CONICYT 2018 Research Fellow (32RF180005). This research has been supported in part by the Gemini Observatory, which is operated by the Association of Universities for Research in Astronomy, Inc., on behalf of the international Gemini partnership of Argentina, Brazil, Canada, Chile, the Republic of Korea, and the United States of America.
Based in part on observations collected at the European Southern Observatory Very Large Telescope in programme 177.D-3023(C) and 179.B-2002. This work presents partial results from the European Space Agency (ESA) space mission {\it Gaia}. {\it Gaia} data are being processed by the {\it Gaia} Data Processing and Analysis Consortium (DPAC). Funding for the DPAC is provided by national institutions, in particular the institutions participating in the {\it Gaia} MultiLateral Agreement (MLA).  This work is based in part on observations made with the Spitzer Space Telescope, which is operated by the Jet Propulsion Laboratory, California Institute of Technology under a contract with NASA. 
\bibliography{1}{}

\bsp

\label{lastpage}

\end{document}